\documentstyle[aps,prl,epsf,floats]{revtex}

\def\Slash#1{{#1\!\!\!\slash}}

\def\nslash{n\!\!\!\slash}
\def\bnslash{\bar n\!\!\!\slash}

\def\OMIT#1{}
\newcommand{\nn}{\nonumber} 

\newcommand{\bn}{\bar n}
\newcommand{\bea}{\begin{eqnarray}}
\newcommand{\eea}{\end{eqnarray}}

\newcommand{\bnP}{\bar {\cal P}}
\newcommand{\bnPp}{\bar {\cal P}_{\!\!+}}

\newcommand{\mcdot}{\mbox{$\cdot$}}
\newcommand{\msim}{\!\hspace{0.1cm}\mbox{$\sim$}\hspace{0.1cm}\!}
\newcommand{\tree}{\mbox{\scriptsize tree}}

\begin{document}
\twocolumn[\hsize\textwidth\columnwidth\hsize\csname@twocolumnfalse\endcsname

\preprint{\tighten \vbox{\hbox{UCSD-01-11} }}

\title{A Proof of Factorization for $B \to D \pi$ }

\author{Christian W. Bauer, Dan Pirjol, and Iain W. Stewart \\[20pt]}

\address{\tighten 
Physics Department, University of California at San
Diego, La Jolla, CA 92093}

\maketitle

{\tighten
\begin{abstract} 

We prove that the matrix elements of four fermion operators mediating the decays
$\bar B^0 \to D^+ \pi^-$ and $B^- \to D^0 \pi^-$ factor into the product of a
form factor describing the $B \to D$ transition and a convolution of a short
distance coefficient with the non-perturbative pion light-cone wave
function. This is shown to all orders in $\alpha_s$, with corrections suppressed
by factors of $1/m_b$, $1/m_c$, and $1/E_\pi$. It is not necessary to assume
that the pion state is dominated by the $q\bar q$ Fock state.

\end{abstract}

}%end tighten
\vspace{0.33in}
]\narrowtext

1.~{\em Introduction.} Our understanding of exclusive $B$ meson decays is
complicated by the non-perturbative nature of the strong interaction. Although
the underlying weak decays are well understood, the hadronic matrix elements are
generally not calculable from first principles. For semileptonic decays these
matrix elements can be parametrized in terms of form factors, which can be
extracted from experiment or lattice simulations. However, for non-leptonic
decays matrix elements of four quark operators are needed, and often little
model independent predictive power can be achieved.

In this letter we present an all orders proof of factorization for $\bar B^0 \to
D^+ \pi^-$ and $B^- \to D^0 \pi^-$, in the limit where the heavy quark masses
approach infinity. To be explicit, we distinguish three types of factorization.
In this letter we prove the generalized factorization of matrix elements of four
quark operators into a form factor describing the $B \to D$ transition and a
convolution of a short distance coefficient with the pion
wavefunction\cite{polwise,bbns}. A second type of factorization, between hard
and infrared scales, is related to defining the correct effective theory as
explained below. Finally, a third type is factorization theorems between soft
and collinear degrees of freedom\cite{sglue} which are also discussed.

The $B \to D \pi$ decays are mediated by the ``full theory'' weak Hamiltonian at
a scale $\mu_0\sim m_b$
\begin{eqnarray}
  {\cal H}_W\! = \frac{4G_F}{\sqrt{2}} V_{ud}^* V^{\phantom{*}}_{cb} \Big[
  C_{\bf 0}^{\rm F}(\mu_0) O_{\bf 0}(\mu_0) \!+\! C_{\bf 8}^{\rm F}(\mu_0)
  O_{\bf 8}(\mu_0) \Big].
\end{eqnarray}
The operators are 
\begin{eqnarray} \label{Oweak}
  O_{\bf 0} &=& \big[\bar{c} \,\gamma^\mu P_L b\big] \big[\bar{d} \,
   \gamma_\mu  P_L u\big]\,,\nn \\
  O_{\bf 8} &=& \big[\bar{c} \,\gamma^\mu P_L T^a b\big] \big[\bar{d} \,
   \gamma_\mu P_L T^a u \big]\,,
\end{eqnarray}
with $P_L = (1-\gamma^5)/2$.  Generalized factorization~\cite{polwise,bbns} says
that for $B \to D \pi$ decays where the light degrees of freedom in the $B$ can
end up in the $D$, the matrix elements of $O_{\bf 0,8}$ can be factored
according to
\begin{eqnarray} \label{ffactor}
 \langle D\pi | O | B \rangle =  {N}\: F^{B\to D}(0)\  \!\int_0^1 
 \!\!\!\!dx\, T(x,\mu)\: \phi_\pi(x,\mu)\,,
\end{eqnarray}
where $F^{B\to D}(0)$ is a ${B\to D}$ form factor at $q^2=0$, $N= i m_B E_\pi
f_\pi/2$, and $\phi_\pi(x,\mu)$ is the non-perturbative light-cone pion
wavefunction~\cite{BL}. Finally, $T(x,\mu)$ is a computable short distance
coefficient and is a function of the renormalization scale $\mu$, the matching
scale $\mu_0$, as well as $x$ and $z=m_c/m_b$.  The earliest form
of~(\ref{ffactor}) is so-called naive-factorization where one sets
$T(x,m_b)\!=\!1$, dropping $\alpha_s(m_b)$ corrections. The first argument for
naive-factorization was based on the idea of color
transparency~\cite{transparency}. The physical picture is simply that long
wavelength gluons cannot resolve the existence of individual colored objects in
the fast moving pion, and thus decouple. The first attempt to prove naive
factorization was by Dugan and Grinstein in the context of a large energy
effective theory (LEET)~\cite{leet}. This theory contains soft gluons coupling
to collinear quarks, but in $n \cdot A = 0$ gauge this coupling vanishes. Thus,
no soft gluons can connect the heavy quarks to the light fermions. However, LEET
omits collinear gluons.  In~\cite{bbns2} the generalized factorization formula
in~(\ref{ffactor}) was shown to be valid at two loops in perturbation theory,
including collinear gluon interactions. This two-loop convolution was reproduced
in a soft-collinear effective theory in~\cite{cbis}. In this paper this theory
combined with heavy quark effective theory (HQET) is used to extend the proof of
factorization to all orders in perturbation theory.

2.~{\em Effective Theory.} The soft-collinear effective
theory~\cite{bfl,bfps,cbis} describes processes with final state particles
having energy much larger than their mass. For $B\to D\pi$ the pion has large
energy, and we take the limit $Q \gg \Lambda_{\rm QCD}$ where $Q$ is $E_{\pi}$,
$m_{b}$, or $m_{c}$.  Momenta $k^\mu \gtrsim Q$ are integrated out and
contribute to Wilson coefficients in the effective theory.  The remaining
infrared physics can be described by including all onshell degrees of freedom
whose momenta are set by the scales in the process.  The heavy mesons can be
described by heavy HQET quarks ($h_v$), soft quarks ($q_s$), and soft gluons
($A_s^\mu$), all with momenta of order $Q\lambda$, where $\lambda\msim
\Lambda_{\rm QCD}/Q \ll 1$.  The fast moving pion contains collinear quarks
($\xi_{n,p}$) and collinear gluons ($A^\mu_{n,p}$), with momenta scaling as
$(p^+, p^-, p^\perp) \sim Q (\lambda^2, 1, \lambda)$. All four components of
$A_{n,q}^\mu$ give order $\lambda^0$ interactions with collinear quarks and are
responsible for binding the pion constituents. In addition, ultrasoft gluons
($A_{us}^\mu$) with momenta $k^\mu_{us} \sim Q\lambda^2$ can be emitted by a
collinear quark without changing the scaling of its momenta (i.e. taking it off
its mass shell).  The HQET fields are labelled by the heavy quark velocity $v$,
while collinear quarks and gluons are labelled by their light cone direction $n$
and the large part of their momentum. The same modes for gluons and quarks also
appear in the method of regions \cite{regions,bbns2}.
\begin{figure}[!t]
 \centerline{\mbox{\epsfysize=4.5truecm \hbox{\epsfbox{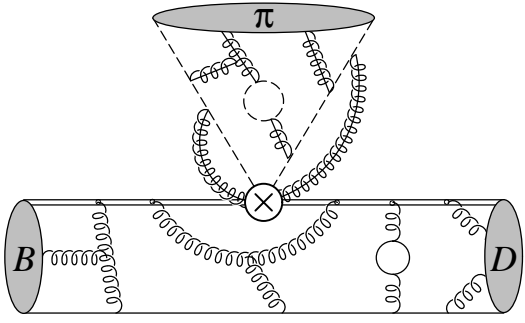}} }} 
\vskip 0.2cm
{\tighten \caption[1]{How the factorization of modes takes place.}
\label{fig:factor} }
\vskip -0.2cm
\end{figure}

To simplify the power counting, fields are rescaled by powers of $\lambda$ to
make all kinetic terms ${\cal O}(\lambda^0)$. This gives $h_v\msim q_s \msim
\lambda^{3/2}$, $(A^+_{n,q},A^-_{n,q},A^\perp_{n,q})\msim
(\lambda^2,1,\lambda)$, $\xi_{n,p}\msim \lambda$, $A_{s}^\mu\msim
\lambda$, $A_{us}^\mu\msim \lambda^2$, and $q_{us} \msim \lambda^3$. Using
topological identities the power of $\lambda$ for an arbitrary diagram can then
be determined entirely from the interaction vertices, and only ${\cal
O}(\lambda^0)$ Feynman rules are required. For $B\to D\pi$ a graph is ${\cal
O}(\lambda^\delta)$ with $\delta +1= \sum_k [(k-8)V_k^{us} + (k-4)
(V_k^s+V_k^{c}+V_k^{cs})]$. $V_k^i$ counts interaction field operators of type
$i$ with scaling $\lambda^k$ ($V^{cs}$ are mixed collinear-soft vertices). For
example, a single $\bar h_{v'} h_v \bar\xi_{n,p'} \xi_{n,p}$ is $V_5^{sc}=1$, so
$\delta=0$. The couplings of soft gluons to heavy quarks are identical to HQET,
and those of soft quarks and gluons are simply given by QCD.

3. {\em Preliminaries.}  We wish to show that at leading order in $\lambda$, the
effective theory Feynman rules only leave diagrams of the form shown in
Fig.~\ref{fig:factor}, so that no non-factorizable infrared contributions occur.
This picture illustrates how, even in the presence of arbitrary hard
interactions, soft gluons decouple from the pion and collinear gluons couple to
the hard vertex (which gives rise to the convolution in~(\ref{ffactor})).
Arguments for the former are fairly standard but are given for our case. The
convolution is more interesting. We begin by showing that in the absence of hard
gluons, collinear gluons completely decouple from the $B$ and $D$ (naive
factorization). We then prove~(\ref{ffactor}) (generalized factorization) by
using the fact that the form of operators induced by integrating out hard gluons
are constrained by a symmetry~\cite{bfps,cbis}.

We will assume that the tail end of wavefunctions are suppressed by $\lambda^a$
with $a>0$. For the pion, these configurations have a single valence quark
carrying off most of the energy, and for the $B$ and $D$ they contain a
spectator with momentum $\gg \Lambda_{\rm QCD}$.  These assumptions can be used
to show the power suppression of annihilation and hard spectator contributions,
respectively~\cite{bbns2}.

4. {\em Naive Factorization.}  To build some intuition, we begin by neglecting
all hard matching corrections proportional to $\alpha_s(Q)$, but work to all
orders in the couplings of the effective theory gluons. In this case we show
that the sum of all diagrams with gluons connecting quarks in the heavy mesons
to those in the pion is zero.

\begin{figure}[!t]
 \centerline{\mbox{\epsfysize=2.2truecm \hbox{\epsfbox{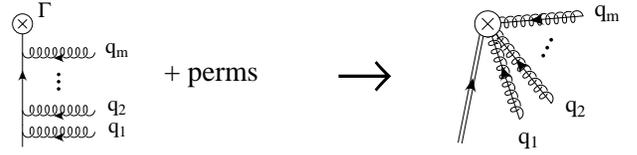}} }} {\tighten
\caption[1]{Matching for the order $\lambda^0$ Feynman rule with a heavy quark
and $m$ collinear gluons. }\label{fig_current} }
\vskip -0.2cm
\end{figure}
Collinear gluons can not couple to the heavy quarks since an HQET quark can not
emit or absorb a collinear gluon and stay near its mass shell~\cite{bfl}.
Instead, the coupling of collinear gluons to heavy quarks introduce non-local
operators, which a priori can still spoil factorization.  To match onto these
operators at tree level we follow~\cite{bfps}. An infinite number of
$A_{n,q}$ gluons contribute to the matching onto any operator with a heavy
quark as in Fig.~\ref{fig_current}.  Since $A_{n,q}^\mu = \bn\mcdot A_{n,q}
n^\mu/2 +{\cal O}(\lambda)$, only the $\bn\cdot A_{n,q}$ gluons appear at ${\cal
O}(\lambda^0)$. For one such gluon
\begin{eqnarray}\label{onegluon}
 && -g \Gamma \frac{m_b \Slash{v} + \Slash{q}_1+m_b}{(m_b v+q_1)^2-m_b^2}
  \frac{\Slash{n}}{2} {(\bar n\cdot A)} \,b \\ 
 && \quad = -g \frac{(\bar n\cdot A_{n,q_1})}{\bn \cdot q_1} \,\Gamma\,
  \frac{(1+\Slash{v}) \Slash{n}}{2 v\cdot n} \,h_v = -g \frac{(\bar n\cdot
  A_{n,q_1})}{\bn \cdot q_1} \,\Gamma \, h_v\,,\nn
\end{eqnarray}
using $\Slash{v} h_v = h_v$. It is important to note that (\ref{onegluon}) is
independent of the value of $v \cdot n$, and thus independent of the heavy
quark velocity $v$. This matching can be extended to include an arbitrary number
of collinear gluons \cite{bfps}
\begin{eqnarray} \label{Wdef}
&&\sum_{m, {\rm perms}} \!\!\!\!
  \frac{(-g)^m}{m!}\: {(\bar n\mcdot A_{n, q_m})\cdots
  (\bar n\mcdot A_{n, q_1}) \over
  (\bn\mcdot q_1)\cdots (\sum_{i=1}^m \bn\mcdot q_i) }\:
  \Gamma \, h_v  \equiv W \,\Gamma\, h_v \,. 
\end{eqnarray}
With these definitions, the effective Hamiltonian below $\mu_0 \sim Q$ matches
at tree level onto the operators
\begin{eqnarray}\label{treeops}
   Q_{{\bf 0},\tree}^{1,5}  &=& \Big[\bar h_{v'}^{(c)}\,\Gamma_h^{1,5}\,
   h_v^{(b)}\Big] \Big[\bar\xi_{n,p'}^{(d)}\, \Gamma_\ell\, 
   \xi_{n,p}^{(u)} \Big]\,, \\ 
  Q^{1,5}_{{\bf 8},\tree}  &=&  \Big[\bar h_{v'}^{(c)}\, \Gamma_h^{1,5}
   (W^\dagger T^A W)\: h_v^{(b)} \Big] \Big[\bar\xi_{n,p'}^{(d)} \,
   \Gamma_\ell \, T^A \xi_{n,p}^{(u)} \Big] \,,\nn
\end{eqnarray}
where $\Gamma_h^{1}= {\nslash}/{2}$, $\Gamma_h^{5}= {\nslash\gamma_5}/{2}$, and
$\Gamma_\ell= {\bnslash(1-\gamma_5)}/{4}$. For $Q^{i}_{{\bf 0},\tree}$ we have
used $W^\dagger W = 1$, which encodes the important observation that collinear
gluon interactions from the $b$ and $c$ quarks cancel identically to all orders
for the color singlet operators.  It is not possible to add additional fields
to~(\ref{treeops}), such as a soft gluon, without increasing the power of
$\lambda$.  A collinear gluon could also interact with the spectator quark in
the $B$ to change it into a collinear quark.  However, this interaction does not
occur at ${\cal O}(\lambda^0)$ because $\Slash{n} \xi_{n,p}=0$.

We defer to the next section the proof that only soft gluons exchanged between
the partons in the $B$ and $D$ contribute, as in Fig.~\ref{fig:factor}. Assuming
this, naive factorization is obtained by showing that $\langle D\pi|$ $Q_{\bf
8,\tree}^{i}|B \rangle$ vanishes, while $\langle Q_{\bf 0,\tree}^{i}\rangle$
factors into the product of matrix elements of two currents.  Let $M$ denote an
arbitrary color structure associated with soft modes exchanged between color
singlet $B$ and $D$ states. Since all adjoint indices in $M$ are contracted, the
lower color trace in $\langle  Q_{{\bf 8},\tree}^i\,  \rangle $ is
\begin{eqnarray} \label{ctrace}
&& {\rm Tr} \big[ M\, W^\dagger T^A W \big] 
 =M\: {\rm Tr}\big[ W^\dagger T^A W \big]
 \propto {\rm Tr}\big[ T^A \big] =0\,.
\end{eqnarray}
By parity $\langle Q_{\bf 0,\tree}^5 \rangle$ vanishes. Finally, $Q_{\bf
0,\tree}^i$ contains no collinear gluons, so {\em no} gluons connect the soft and
collinear partons at ${\cal O}(\lambda^0)$. Thus, $\langle Q_{\bf 0,\tree}^i
\rangle$ factors
\begin{eqnarray} \label{nfact}
 \big\langle D_{v'}\,\pi_n \big|\, Q_{{\bf 0},\tree}^{1} 
 \big| B_v \big\rangle &=& \frac{i}{2}\: E_\pi f_\pi \, m_B  F^{B\to D}(0)\: 
  +  \ldots \,.
\end{eqnarray}
Eq.~(\ref{nfact}) is the product of the pion decay constant from $E_\pi
f_\pi=\frac{i}{2}\langle \pi_n | \bar\xi_{n,p'}\, \Slash{\bn}\,\gamma^5\,
\xi_{n,p}\, | 0 \rangle$ with $p_\pi^\mu=E_\pi n^\mu$, and the $B\to D$ form
factor $F^{B\to D}(0) = \frac12 (m_D/m_B)^{1/2} (1\!+\!m_B/m_D)\,\xi(v\mcdot
v')$, where $\xi(v\mcdot v')$ is the Isgur-Wise function~\cite{bbook}. The
states in~(\ref{nfact}) are in the effective theory (with relativistic
normalization), and the ellipses denote terms suppressed by $1/Q$ or
$\alpha_s(Q)$.  The result in~(\ref{nfact}) is exactly the statement of naive
factorization.

\widetext
\begin{figure}[!t]
\begin{eqnarray}
 \begin{picture}(270,50)(100,0)
 \mbox{\epsfxsize=16.truecm \hbox{\epsfbox{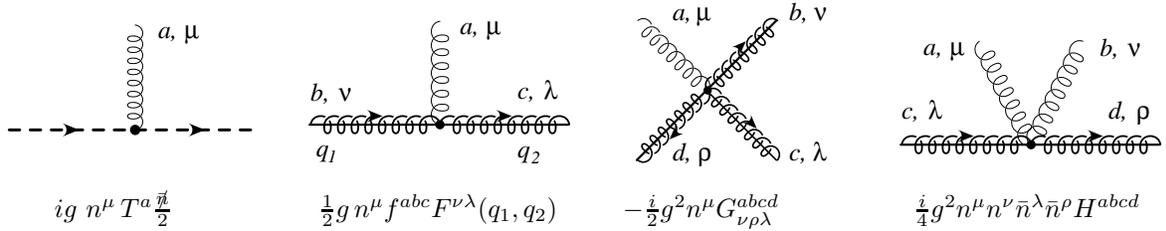}}  }
 \put(-430,-15){$\mbox{\normalsize $i g\,$} \, n^\mu \, T^a
  \frac{\bar\nslash}{2}$}  
  \put(-330,-15){$\frac{1}2 g\, n^\mu f^{abc}  F^{\nu\lambda}(q_1,q_2)$}
  \put(-215,-15){$-\frac{i}{2}{g^2} n^{\mu} G^{abcd}_{\nu\rho\lambda}$}
  \put(-105,-15){$\frac{i}{4} g^2  n^{\mu}n^{\nu} \bn^{\lambda}
  \bn^{\rho}  H^{abcd}$}
  \end{picture} \nn
\end{eqnarray}
\medskip {\caption[1]{Order $\lambda^0$ Feynman rules for coupling ultrasoft or
soft gluons (spring lines) to ``collinear'' fermions (thick dashed lines) and
``collinear'' gluons (thick spring+solid lines) (c.f. section 5).  As an
example, coupling an ultrasoft gluon to two collinear gluons in background field
Feynman gauge one finds $ F^{\nu\lambda}(q_1,q_2) = 2 \bn \mcdot q_1\,
g^{\nu\lambda}$, which is $V_4^c=1$.  
\OMIT{$G^{abcd}_{\nu\rho\lambda} = $
$f^{abe}f^{cde}(g^{\nu\rho}\bn^\sigma\!-g^{\nu\sigma}\bn^\rho)$ $\!+
f^{ace}f^{bde}(g^{\rho\sigma}\bn^\nu-g^{\nu\sigma}\bn^\rho)$ $\!+
f^{ade}f^{bce}(g^{\rho\sigma}\bn^\nu-g^{\nu\rho}\bn^\sigma)$ and
$H^{abcd}=f^{ade}f^{bce}\!+ f^{ace}f^{bde}$.}  
}\label{fr1} }
\vskip -0.25cm
\end{figure} %\vspace{6.2cm}\phantom{x}\vspace{6.2cm}\hspace{-.5cm} 
\narrowtext 

5. {\em Decoupling of Ultrasoft and Soft Gluons.}  By simple power counting the
couplings of ultrasoft gluons to heavy quarks and soft modes are suppressed by
at least one power of $\lambda$. For e.g., $\bar h_v A_{us}^\mu h_v\msim
\lambda$, i.e.\ $V_5^s=1$. (If ultrasoft heavy quarks are allowed as
in~\cite{bbns2} then decoupling $A_{us}^\mu$ gluons follows the proof for
$A_s^\mu$ gluons below.) To prove factorization, we therefore need to show that
interactions between soft gluons and ``collinear'' particles with $\bn\mcdot p
\msim Q$ decouple.  For this section only, the name ``collinear'' will be used
to refer to {\em any} particles with $\bn\mcdot p \msim Q$. This includes the
degrees of freedom discussed in section 3, as well as offshell fluctuations with
$p^2\! <\!  Q^2$ (for example quarks and gluons with momenta
$(k^+,k^-,k^\perp)\msim Q(\lambda,1,\lambda)$).

The decoupling of soft gluons from collinear particles is a standard part of the
proof of QCD factorization theorems for processes such as
Drell-Yan~\cite{sglue}.  The decoupling depends on only soft $n\mcdot { A}_s$
gluons coupling to collinear fields at ${\cal O}(\lambda^0)$, and that the soft
$k^-$ and $k^\perp$ momenta

\phantom{x}\vspace{4.25cm}\hspace{-.5cm}
\noindent
%\indent 
drop out of collinear propagators. Applying Ward identities then factors
arbitrary soft attachments out of any time ordered product of collinear fields.

The power counting can be used to derive that only $n\mcdot {A}$ soft
(ultrasoft) gluons couple to collinear particles at $\lambda^0$.  (The offshell
collinear modes can be included by treating them as auxiliary fields.)  At
lowest order we find only the Feynman rules shown in Fig.~\ref{fr1}.
We see immediately that all soft (ultrasoft) gluons couple proportional to
$n^\mu$.

In the effective theory the soft $k^-$ and $k^\perp$ momenta drop out of
collinear propagators.  This occurs due to the large $\bn\mcdot p_c$ component
for a collinear momentum $p_c$, so that $(p_c\!+\!  k)^2=\bn\mcdot p_c\: n\mcdot
k+{\cal O}(\lambda^2)$. For ultrasoft gluons these momenta drop out using the
multipole expansion and equations of motion in the Lagrangian~\cite{bfl}.

Now, $n\mcdot {A}_s$ gluons couple to a collinear time ordered product $T_c$
which has dependence only on $k^+$ momenta, for e.g.~${A}_s(k)\mcdot T_c$ $=
n\mcdot {A}_s\, \bn\mcdot T_c(k^+)/2$ = $ n\mcdot {A}_s\: k\mcdot T_c/(n\mcdot
k)$. Thus, QCD Ward identities can be applied.  By induction all soft gluons can
be decoupled from $T_c$ into eikonal line prefactors~\cite{sglue},
$S=P\,\exp[ig\int \! dx\: n\mcdot {A}_s(x n^\mu)]$.  For the operator $Q_{\bf
0}^i$, unitarity gives $S^\dagger S=1$ and the soft gluons decouple.  For
$Q_{\bf 8}^i$ one obtains a color structure $T^a\otimes W S^\dagger T^a S
W^\dagger = S T^a S^\dagger \otimes W T^a W^\dagger$ and the vanishing of the
octet matrix element in~(\ref{ctrace}) is still obtained.

6. {\em Generalized Factorization.}  To include arbitrary hard corrections we
can not rely on tree level matching as was done to determine the operators
in~(\ref{treeops}). Since momenta $\ge Q$ are integrated out, the Wilson
coefficients in the effective theory are in general arbitrary functions of the
large $\bn\cdot p_i$ momenta~\cite{bfps}. In~\cite{cbis} it was pointed out that
this functional dependence is greatly restricted by a symmetry induced by
collinear gauge transformations. Under this symmetry, $\xi_{n,p}$ and
$A_{n,q}^\mu$ transform, but $h_v$ does not since collinear gluons do not couple
to nearly onshell heavy quarks. For $B \to D \pi$ the most general allowed
leading order operators are~\cite{cbis}
\begin{eqnarray}\label{realQs}
  Q^{j}_{\bf 0} \!&=&\!  \Big[\bar h_{v'}^{(c)} \,\Gamma_h^{j}\, 
   h_v^{(b)}\Big] \Big[\bar\xi_{n,p'}^{(d)} W C^{j}_{\bf 0}(\bnPp)\,
    \Gamma_\ell\,  W^\dagger \: \xi_{n,p}^{(u)}\Big], \\ 
  Q^{j}_{\bf 8} \!&=&\!  \Big[\bar h_{v'}^{(c)} S \Gamma_h^{j} T^a S^\dagger 
    h_v^{(b)}\Big]\! \Big[\bar\xi_{n,p'}^{(d)} W  C^{j}_{\bf 8}(\bnPp) 
    \Gamma_\ell T^a W^\dagger \xi_{n,p}^{(u)}\Big] .\nn
\end{eqnarray}
where $j=1,5$.  Helicity ensures that only $\Gamma_\ell$ is needed between the
light quarks.  The dimensionless Wilson coefficients $C^i_{\bf a}$ are functions
of the renormalization scale $\mu$, as well as $m_b$, $m_c$, $v\mcdot v'$, and
the label operators $\bnP$ and $\bnP^\dagger$~\cite{cbis}.  Since $\bnP$ does
not commute with collinear fields the short distance Wilson coefficient is
conveniently included as part of the $Q_{\bf a}^i$'s. In terms of the label
operators
\begin{eqnarray} \label{W}
  W &=& \Big[ \lower7pt \hbox{ $\stackrel{\sum}{\mbox{\scriptsize perms}}$ }
  \!\! \exp\Big( \!-\!g\,\frac{1}{\bnP}\ \bn\mcdot A_{n,q} \Big) \Big] \,.
\end{eqnarray}
$W^\dagger \xi_{n,p}$ is an invariant under a collinear gauge transformation.
The operators $\bnP$ and $\bnP^\dagger$ give the sum of labels on collinear
fields to their right and left respectively, and are described in detail
in~\cite{cbis}. For e.g., if $f$ is some function then $f(\bnP) \big(
\bar\xi_{n,p'}\, A^\mu_{n,q}\, A^\nu_{n,r}\, \xi_{n,p} \big) = f(\bn\mcdot
q\!+\! \bn\mcdot r\!+\!\bn\mcdot p\!-\!\bn\mcdot p') \\* \times\! (
\bar\xi_{n,p'} A^\mu_{n,q} A^\nu_{n,r} \xi_{n,p} )$. For the $B\to D\pi$
matrix element the combination $\bnP^\dagger-\bnP$ behaves like a total
derivative, and by momentum conservation gives the total large momentum label of
the effective theory state\cite{cbis}, $\bnP^\dagger-\bnP=2E_\pi$. The dependence
on the other linear combination $\bnPp= \bnP + \bnP^\dagger$ is displayed
explicitly in~(\ref{realQs}).

If we neglect hard corrections ($C_{\bf 0}^i=1$) and leave soft-collinear
couplings in a Lagrangian then~(\ref{realQs}) reduces to~(\ref{treeops}). This
follows from the color identity $W^\dagger T^A W \otimes T^A = T^A \otimes W T^A
W^\dagger$, which connects the picture where $W$ is obtained by integrating out
offshell heavy quarks to the picture where $W$ appears by demanding invariance
under collinear gauge symmetry in the effective theory.
 
Up to power corrections the full theory matrix element is $\langle D\pi|
C^F_{\bf a} O_{\bf a}| B\rangle = \langle D_{v'}\pi_n| Q_{\bf b}|B_v\rangle$
(summing over ${\bf a,b}$).  Therefore, we must simply prove generalized
factorization for the effective theory matrix element. For $B\to D\pi$ the same
arguments used in section 4 rule out contributions from $Q_{\bf 8}^i$ and
$Q_{\bf 0}^5$. For $Q_{\bf 0}^1$,
%\begin{mathletters}
\begin{eqnarray} \label{step1}
  && \big\langle D_{v'}\pi_{n} \big| Q^1_{\bf 0} \big| B_v \big\rangle \\
  && = m_B  F^{B\to D}\
      \big\langle \pi_n \big| \bar\xi_{n,p'} \Gamma_\ell
     \,W\, C_{\bf 0}^1(\bnPp) \, W^\dagger \: \xi_{n,p}\big| 0\big\rangle \nn\\
  && = m_B  F^{B\to D}\!\!\! \int \!\!\! d\omega\,
    C_{\bf 0}^1(\omega) 
    \big\langle \pi_n \big| \bar\xi_{n,p'} \Gamma_\ell \,W 
   \delta(\omega\!-\!\bnPp) W^\dagger  \xi_{n,p}\big| 0\big\rangle . \nn
\end{eqnarray} 
%\end{mathletters} \\[-24pt]
%\noindent 
In the first equality we used that collinear gluons do not connect to particles
in the heavy meson states, while soft gluons do not connect to those in the
pion. The second equality follows trivially, but illustrates how the
non-commutative nature of the Wilson coefficients and fields leads to a
convolution.  In our formulae hard corrections to the ${B\to D}$ form factor are
contained in $C_{\bf 0}^1$.

Next we show that the matrix element in the last line of~(\ref{step1}) is the
Fourier transform (FT) of 
\begin{eqnarray} \label{step3}
 &&  \big\langle \pi^-_n(p_\pi) \big| \bar\xi_{n}^{(d)}(y)\, \bnslash 
   \gamma_5\, {\cal W}(y,-y)\, \xi_{n}^{(u)}(-y) \big| 0 \big\rangle 
  \nn \\
 && \quad 
    \equiv -2 i f_\pi E_\pi\! \int_0^1\!\!\! dx\, \phi_\pi(x,\mu)\, 
    e^{i 2 y  E_\pi (2x-1)} \,,
\end{eqnarray} 
where the FT of $\xi_n(y)$ is $\xi_{n,p}$, ${\cal W}(y,-y)$ is the path
ordered eikonal line from positions $-y\bn^\mu$ to $y\bn^\mu$, and
$\phi_\pi(x,\mu)$ is the light-cone pion wavefunction. Since the
FT of $[\,\bar\xi_{n}(y) {\cal W}(y,\infty)]$ with respect to~$R$ is
$\bar\xi_{n,p}W \delta_{\bnP^\dagger\! , R}\,$, the Fourier transform of
(\ref{step3}) is
\begin{eqnarray} \label{step2}
 && \int\!\! \frac{dy}{2\pi} \: e^{i\omega(-y)}\: \big\langle \pi_n\big| 
  \bar\xi_{n}(y) \Gamma_\ell {\cal W}(y,-y) \xi_{n}(-y)  
  \big| 0 \big\rangle \nn \\
 && = \int\!\! \frac{dy}{2\pi}\: e^{-i\omega y}\: \big\langle 
  \bar\xi_{n}(y) {\cal W}(y,\infty) \Gamma_\ell {\cal W}^\dagger(-y,\infty) 
  \xi_{n}(-y)   \big\rangle \nn \\
 && = \int\!\! \frac{dy}{2\pi}\, \sum_{R,T}\: e^{-i\omega y}\,e^{i(R+T)y} 
  \big\langle\bar\xi_{n,p'} \Gamma_\ell W \delta_{\mbox{\scriptsize 
  $\bnP^\dagger$}\!\!,R}\:  \delta_{\bnP,T} W^\dagger \xi_{n,p} \rangle \nn\\
 && = \sum_{R,T}\: \delta(\omega\!-\!T\!-\!R)\: \langle \bar\xi_{n,p'} 
   \Gamma_\ell W \delta_{\mbox{\scriptsize $\bnP^\dagger$}\!\!,R}\: 
   \delta_{\bnP\!,T} W^\dagger \xi_{n,p}\rangle \nn\\
 && = \big\langle \pi_n\big|   \bar\xi_{n,p'} \Gamma_\ell\,
   W\, \delta(\omega\!-\bnPp)\, W^\dagger\, \xi_{n,p} \big| 0 \big\rangle \,.
\end{eqnarray}
Thus, the convolution in Eq.~(\ref{step1}) is
\begin{eqnarray} \label{step4}
 && \int \!\! d\omega\, C_{\bf 0}^1(\mu,\omega) \big\langle \pi_n 
   \big| \bar\xi_{n,p'} \Gamma_\ell \,W 
   \delta(\omega\!-\!\bnPp) W^\dagger \, \xi_{n,p}\big| 0\big\rangle \nn \\
 && = \frac{i}{2} f_\pi E_\pi \! \int\!\!  \frac{dy}{2\pi}\: d\omega\! 
  \int_0^1\!\!\!\! dx\, e^{iy [2(2x-1)E_\pi-\omega]} \, 
  C_{\bf 0}^1(\mu,\omega)\, \phi_\pi(x,\mu) \nn\\
 && = \frac{i}{2} f_\pi E_\pi \!\int_0^1\!\!\!\! dx\, C_{\bf 0}^1\big(\mu,
   2(2x-1)E_\pi\big)\: \phi_\pi(x,\mu) \nn\\
 && = \frac{i}{2} f_\pi E_\pi \!\int_0^1\!\!\!\! dx\: T(x,\mu)\: \phi_\pi(x,\mu) 
  \,,
\end{eqnarray}
where $T(x,\mu)\equiv C_{\bf 0}^1(\mu,(4x-2)E_\pi)$. 

Combining (\ref{step1}) and (\ref{step4}) we arrive at 
\begin{eqnarray} \label{result}
 && \big\langle D_{v'}\pi_{n} \big| Q^1_{\bf 0} \big| B_v \big\rangle 
%\\ &&\ 
 = N  F^{B\to D}(0)\,  \!\int_0^1\!\!\!\! dx\, T(x,\mu)\,
  \phi_\pi(x,\mu) \,.\nn %\\[-7pt] \\[-20pt]\nn
\end{eqnarray}
This is our final result, and it reproduces the generalized factorization
formula in Eq.~(\ref{ffactor}).  Note that it was not necessary to set the
transverse momenta of partons to zero. It should be fairly obvious that
this proof also goes through for other class I decays. Q.E.D. \\[-9pt]

This work was supported by the DOE under grant DOE-FG03-97ER40546 and by NSERC
of Canada.

\vspace{-0.5cm}

{\tighten

}

\end{document}